# EMEEDP: Enhanced Multi-hop Energy Efficient Distributed Protocol for Heterogeneous Wireless Sensor Network

Sunil Kumar[1], Dr. Priya Ranjan[2], Dr. R. Radhakrishnan[3]
Department of CSE[1&3], Department of EEE[2]
ABES Engineering College Ghaziabad[1&3], Amity University, Noida[2]
Sunil.kumar@abes.ac.in[1], pranajan@amity.in[2], radhakrishnanramaswami@gmail.com[3]

Abstract— In WSN (Wireless Sensor Network) every sensor node sensed the data and transmit it to the CH (Cluster head) or BS (Base Station). Sensors are randomly deployed in unreachable areas, where battery replacement or battery charge is not possible. For this reason, Energy conservation is the important design goal while developing a routing and distributed protocol to increase the lifetime of WSN.

In this paper, an enhanced energy efficient distributed protocol for heterogeneous WSN have been reported. EMEEDP is proposed for heterogeneous WSN to increase the lifetime of the network. An efficient algorithm is proposed in the form of flowchart and based on various clustering equation proved that the proposed work accomplishes longer lifetime with improved QOS parameters parallel to MEEP.

A WSN implemented and tested using Raspberry Pi devices as a base station, temperature sensors as a node and xively.com as a cloud. Users use data for decision purpose or business purposes from xively.com using internet.

## I. INTRODUCTION

WSN is composed of various low powered, low storage and short range communication devices/nodes. Each node is capable of sensing the region in its range and transmit it to its neighbour's node or cluster head using wireless links. A Sensor node is composed of power, sensing, processing and communication unit typically. Deployment of sensors are random and static, only few sensors are dynamic [1].

In homogeneous WSN, sensors would have the same lifetime if they have the same energy consumption rate. In heterogeneous WSN, each sensors have different capabilities in terms of storage, processing, sensing, and communication. In Computational heterogeneity a heterogeneous node has a more powerful microprocessor and large memory compare to normal node. In Link heterogeneity a heterogeneous node has high bandwith and longer transmission range and provide more secure & reliable data transmission. In Energy heterogeneity a heterogeneous node has different energy level or it may be line powered, or its battery is replaceable. These types of heterogeneous nodes called advanced nodes.

Using Heterogeneous WSN, network lifetime, reliability of data transmission will increase while latency of data transportation will decrease. To evaluate the performance of heterogeneous WSN we considered network lifetime, number of cluster heads per round, number of alive (total, super, advanced and normal) nodes per round, and throughput as evaluation parameters.

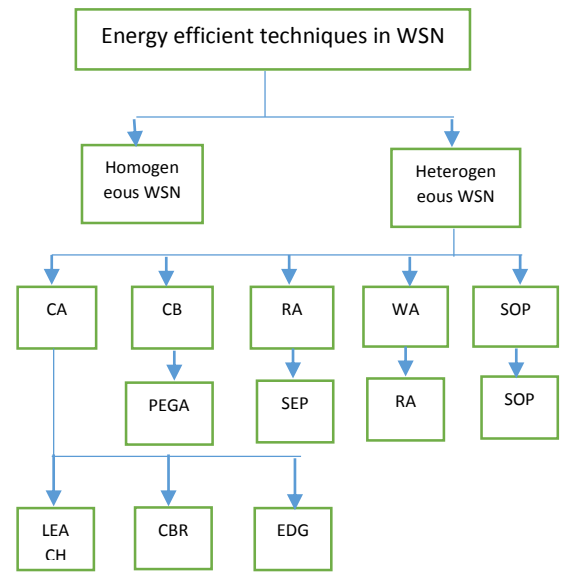

*Fig. 1 Energy efficient taxonomy of heterogeneous WSN*

In figure 1 a taxonomy of heterogeneous WSN is given and among of various techniques cluster based approach is best for solving the sensors energy constraints by reducing the cost of data transmission and aggregation of sensed data before transmitting data to base station [2]. In a cluster, there is a cluster head which is selected randomly based on available energy so that the energy load can be evenly distributed in sensor network [2]. All the sensor node sense the data and transmit it to the cluster head and after aggregation it pass the data to the base station. The base station (Raspberry Pi) is a larger computer or mote where data from the sensor network will be aggregated, compiled and processed. The base station may communicate with any cloud services via Internet or Satellite [1] [2]. Human operators controlling the sensor network send commands and receive responses through the base station. Raspberry pi device can be used as base station and forward all the data at cloud (Xively.com site) for decision making or various business purposes.

Clustering using a two layer approach in the WSN where first layer is used for selecting the cluster head and the second layer for routing [1]. With the introduction of heterogeneity in sensor network energy can be further saved without sacrificing performance. Nodes in a WSN are deployed closely therefore multi-hop communication is desirable for achieving the energy efficiency. A node joins a cluster which is very near to it depending upon signal strength. A sleep state has been presented during the cluster formation process which further increases the network life and create energy efficient WSN. Here a multi-hop and clustering scheme are combined together with advanced node capabilities in distributed environment and proposes a new EMEEDP to improve sensor network lifetime with reduced delay. In Section 2 cluster based related protocols are described. In section 3 problem statement is define which is related to energy efficient designing in WSN. Section 4 describes the heterogeneous network model and various equations are defined for energy calculation in clustered WSN, section 5 describes the proposed EMEEDP protocol and finally the paper is concluded in section 6.

## II. RELEATED WORK

Low Energy Adaptive Clustering Hierarchy (LEACH) protocol introduced in 2000 [2]. LEACH is a clustering protocol where cluster heads responsibility are rotated to balance the energy in WSN. The LEACH works on how to define cluster and cluster head (CH). The idea is to form clusters for the sensor nodes based on the received signal strength (RSSI) and use local cluster heads as routers to the base station. CH accepts data from its members, aggregate the data and forward it to the Base Station (BS). All the data processing such as data fusion and aggregation are local to the cluster. All CHs directly communicate to the BS using star topology. Since CH are randomly chooses in LEACH algorithm so there is some probability to form a low-energy normal node as a CH. The node becomes a cluster head for the current round if the number is less than the following threshold:

$T(n) = [p / (1-p \times (r \bmod (1/p)))]$ if $n \in G$ otherwise 0.

Here G denotes the set of nodes that are not selected as a cluster head in last $1/p$ rounds and r is the current round.

Power-Efficient Gathering in Sensor Information Systems (PEGASIS) [4], is a near optimal chain-based protocol. The basic idea of the protocol is that in order to extend network lifetime, nodes need only to communicate with their closest neighbours and they take opportunities in communicating with the base-station. To locate the closest neighbour node in PEGASIS, each node uses the signal strength to measure the distance or location to all neighbouring nodes and then adjust the signal strength so that only one node can be heard. The chain in PEGASIS will consist of those nodes that are closest to each other and form a path to the base-station. The aggregated form of the data will be sent to the base station by any node in the chain and the nodes in the chain will take turns in sending to the base-station.

Hybrid Energy Efficient Distributed clustering Protocol (HEED) protocol proposed in 2004 [5]. It extends the basic scheme of LEACH by using residual energy as primary parameter and network topology features (e.g. node degree, distances to neighbours) as secondary parameters. The clustering process is divided into a number of iterations, and in each iterations, nodes which are not covered by any cluster head double their probability of becoming a cluster head. Since these energy-efficient clustering protocols enable every node to independently and probabilistically decide on its role in the clustered network, they cannot guarantee optimal elected set of cluster heads.

Threshold-sensitive Energy Efficient sensor Network protocol (TEEN) [7][8] is a reactive protocol where sensor nodes sense the medium continuously, but the data transmission is done less frequently. A CH broadcast its members a hard threshold (HT), which is the absolute value of an attribute and a soft threshold, which is a small change in the value of the sensed attribute that triggers the node to switch on its transmitter and transmit. The HT tries to reduce the number of transmissions by allowing the nodes to transmit only when the sensed attribute is in the range of interest. The ST further reduces the number of transmissions when there is little or no change in the sensed attribute. A smaller value of the ST gives a more accurate picture of the network, at the expense of increased energy consumption.

Stable Election Protocol SEP [9] is a cluster based protocol for two level heterogeneous network. In SEP there are two types of nodes: normal and advanced. Advanced nodes have more energy than the normal nodes and it is the source of heterogeneity in the network. Weighted probabilities of normal and advanced nodes are used to determine the thresholds for the election of cluster head in a round.

DEEC [11] [16] is a proactive protocol designed for two and multilevel heterogeneous network where CH are elected by a probability based on the ratio between residual energy of each node and the average energy of the network. The node with higher initial and residual energy has greater probability to become a CH. DEEC as in LEACH rotates the role of CH among all the nodes to expend energy uniformity in WSN.

Elbhiri et al. [10] proposed a developed distributed energy efficient clustering scheme where all nodes use the initial and residual energy level to define the cluster heads.

EEHC [13] is an energy efficient scheme for WSN. By setting powerful node EEHC increases the life of the network by 10 % as compared to LEACH.

EECDA [16] is a cluster based protocol where three level in a heterogeneous networks are used. Some Advanced nodes of the network has more energy compare to normal nodes. Some super nodes has more energy compare to advanced nodes. Novel cluster head election and a path of maximum sum of energy residues for data transmission is used in EECDA for increasing network lifetime and stable region.

DBCP [15] is an energy efficient clustered protocol for heterogeneous wireless sensor network which selects the cluster heads according to initial energy and average distance.

### III. PROBLEM STATEMENT

In wireless sensor network various issues need to be dealt while designing an energy optimization techniques distributed environment. Some are:
- Selection of cluster head.
- Number of active sensors between sensor node and cluster head.
- Distance between cluster head node and gateway node.
- Mobile beacon trajectory path (A node has moving capability with and GPS enabled, used in localization so that other sensors find out their coordinates).
- Cost metrics (Prorogation delay, processing delay, and queuing delay) in communication between sensors.
- Interoperability capability between the sensor nodes (heterogeneous environment).
- Reliability of data transmission or security in WSN.
- latency of data transportation

In this proposed work, we are considering heterogeneous environment to reduce the energy in distributed WSN.

### IV. HETEROGENEOUS NETWORK MODEL

This section defines the WSN model of the proposed protocol. Model contains n number of sensor nodes, randomly deployed in a 100 × 100 square meters region as shown in figure 1. X represent base station, o represent normal node and + represent advanced node in figure 1. Various assumptions of the network model and sensor nodes are as follows.

- Nodes are deployed randomly and uniformly in the sensing region.
- Base station and nodes become stationary after deployment OR base station has limited mobility.
- Nodes are location unaware i.e. they do not have any information about their location. Localization technique can be used for awareness of location/coordinates.
- Nodes continuously sense the region and they always have the data to send the base station.
- Battery of the nodes cannot be changed or recharged due to harsh environment deployment.

Sensing region has two types of sensor nodes i.e. advanced and normal nodes. Let initial energy of normal nodes is E0 and m be the fraction of advanced nodes having initial energy [E0 × (1 +α)], where α means that advanced node have α times more energy than normal node. Heterogeneous network total initial energy is given by equation 1 in table 1.

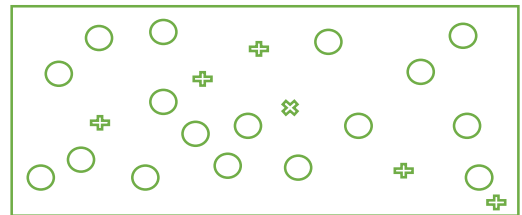

Figure 1 Sensors deployment area

Free space ($d^2$ power loss) and the multipath fading ($d^4$ power loss) channel both techniques are used based on the distance between the transmitter and receiver. When distance is less than specific a threshold value then free space model are used otherwise multipath loss model is used with conditions. The amount of energy required to transmit L bit packet over a distance d is given by Equation 2 in table 1.

### V. PROPOSED EMEEDP

In EMEEDP to calculate energy parameters in heterogeneous WSN a table is given below.

| Eq. No | Contents | Equation |
|---|---|---|
| 1 | Heterogeneous network total initial energy | $ETotal = N \times E0 (1 + \alpha m)$ |
| 2 | Energy required to transmit L bit packet over a distance d, $Eelec$ is the electricity dissipated to run the transmitter or receiver | $ETx(L,) = L \times Eelec + L \times \varepsilon fs \times d^2$ if $(d < d0)$ <br><br> $L \times Eelec + L \times \varepsilon mp \times d^4$ if $(d \geq d0)$ |
| 3 | For receiving an L bit message the energy expends by radio | $ER(L) = L \times Eelec$ |
| 4 | Energy dissipated in the cluster head | $ECH = (n/k - 1) \times L \cdot Eelec + (n/k)L \cdot EDA +$ |
|   | during a single round | $L \cdot Eelec + L \cdot \varepsilon fs \cdot d^2 toBS$ |
| 5 | Energy dissipated in a non-cluster head | $ENCH = L \times (Eelec + \varepsilon fs \times d^2 toCH)$ |
| 6 | Energy consumed in a round | $ECluster = ECH + ((n/k)ENCH)$ |
| 7 | Total energy dissipated in the network | $ET = L \times (2nEelec + nEDA + k\,\varepsilon fs\,d^2 toBS + n\varepsilon fs\,d^2 toCH)$ |
| 8 | Average distance from cluster head to the BS | $dtoBS = 0.765(M/2)$ |
| 9 | Normal and advanced node weighted probability | $pnrm = popt/(1+\alpha \cdot m)$  $padv = [popt/(1+\alpha \cdot m)] \times (1+\alpha)$ |
| 10 | Optimal number of clusters | $kopt = (M/\,d^2\,toBS)$ |
| 11 | Node's optimal probability to become cluster head in round | $popt = (kopt/\,n)$ |
| 12 | Normal and advanced nodes weighted probability | $pnrm = popt/(1+\alpha m)$  $padv = [popt/(1+\alpha \cdot m)] \times (1+\alpha)$ |
| 13 | Average probability $pi$ | $pi = [popt \times Ei\,r]\,/\,[(1+\alpha \cdot m) \times Ei\,t]$  *if si is the normal node*  $pi = popt\,(1+\alpha)/\,(1+\alpha \cdot m)$  *if si the advanced node* |
| 14 | The probability threshold $T(si)$ which node $si$ uses | $T(si) = pi\,/[(1-pi)\,r\,mod\,(1/\,pi)\,]$ *if si ε G'*  $0$ *otherwise* |

*Table 1 Equations used in EMEEDP heterogeneous WSN*

EMEEDP is a clustering protocol for two level heterogeneous sensor networks. EMEEDP consists of n number of normal nodes, advanced nodes and single base station which is placed near the middle of the deployment area. The distance of any sensor node, advanced node to its CH or sink is ≤ d0. The energy dissipated in the cluster head during a single round is given in Equation (4) in table 1. Hence all the sensor nodes transmit data to CH and if any node is not cover by any cluster head then energy dissipated calculated by equation (5) in table 1. We can calculate all the energy parameters as specified in table 1.

Epoch of the network increases in proportion of the energy increment. Heterogeneous nodes increases the system energy by $(\alpha \cdot m)$ times hence for optimizing the stable region, new epoch must become equal to $[1\,/popt\,(1+\alpha \cdot m)]$.

Let $pnrm$, $padv$ denotes the weighted election probability for normal and advanced nodes respectively. The average number of cluster heads per round per epoch is equal to $(n \times (1+\alpha \cdot m) \times pnrm)$. Normal and advanced nodes weighted probability can be calculated by using the equation 12 in table 1.

In a heterogeneous network, to guarantee that there is average $(n \times popt)$ cluster-heads in every round, each node $si$ becomes a cluster-head once every $ni = 1/popt$ rounds. When the network operates for some time then the nodes cannot have the same residual energy. Thus energy is not well distributed if the rotating epoch $ni$ is equal for all nodes. Hence low-energy nodes will die quickly than high-energy nodes of the network. To get rid of this problem EMEEDP protocol chooses different $ni$ for normal nodes on the basis of residual energy $(r)$ of node $si$ in a round r. Let $pi = 1\,/ni$ be average probability to become a cluster-head during $ni$ rounds. When nodes have the same energy at each epoch, choosing the average probability $pi$ to be $popt$ can ensure that there are $n \times popt$ cluster-heads in each round and all nodes die approximately at the same time. When nodes have different energy, $pi$ of the nodes with more energy should be larger than $popt$. Let $Ei(t)$ denotes the initial energy and $Ei(r)$ represents residual energy of a normal node $si$ at round r, using $Ei(t)$ to be the reference energy, for normal nodes we have $popt = [popt(Ei(r)/Ei(t))]$, $popt$ is the reference value of average probability and determines the threshold $T(si)$ of node $si$. In two-level heterogeneous networks, EMEEDP replace the reference value $popt$ with the weighted probabilities given in Equation 11 for normal nodes while advanced node weighted probability will remain the same. Hence $pi$ changed according to equation 13. The probability threshold $(si)$ which node $si$ uses to determine whether it can become a cluster-head in a round can be calculated by using the Equation 14. When $si$ is normal node then $G'$ represents normal node set otherwise $G'$ represents advanced node set that are not elected as cluster heads within $1/\,pnrm$ or $1/\,padv$ rounds.

Another important factor when we are creating CH is that they should not overlap each other hence the distance between them must be defined. Using localization technique or RSSI, selected cluster head will cover all the deployment area.

In a round if a normal sensor node becomes a cluster head then after collecting the data from its members it aggregates the data and instead of sending the data directly to sink it will try to find out an advanced node such that

- ➢ Node is not a cluster head in this round.
- ➢ Distance between normal CH and AN is less than the distance between normal CH and BS.

If the normal cluster head find such an advanced node that is not a cluster head in this round r and also its distance is less than distance between normal CH and BS then normal CH instead of sending the data directly to the BS it sends its data to this AN which further send it to the BS. If normal cluster head does not find any such AN who fulfils the above mentioned two conditions then it will send the aggregated data of its members directly to the BS itself. Thus by introducing a gateway concept or three tier architectures for normal cluster head, EMEEDP has reduced the energy consumed in transmission to prolong the network lifetime and stability period. Moreover, when the CHs are selected, each node joins to the closest (considering the transmission power) CH. However in some cases this is not the optimal choice because, if exists a sensor node in the BS direction whose distance from the BS is less than all the nearby cluster-head distance.

Let us consider the figure 2, where the node N1 has to transmit L-bits message to the BS. The closest CH to N1 is CH1. And, if the node belongs to this cluster, it spends energy.

$E1 = L \cdot Eelec + L \cdot \epsilon d2 \cdot d2^x$

Where $x = 2$, $\epsilon d2 = 10$ pj / bit / m², if $d2 < d0$

$x = 4$, $\epsilon d2 = 0.0013$ pj / bit / m⁴, if $d2 \geq d0$

But if the node N1 chooses to transfer data to the base station directly, this energy will be:

$E2 = L \cdot Eelec + L \cdot \epsilon d3 \cdot d3^y$

Where $y = 2$, $\epsilon d3 = 10$ pj / bit / m², if $d3 < d0$

$y = 4$, $\epsilon d3 = 0.0013$ pj / bit / m⁴, if $d3 \geq d0$

Here positive coefficients $x$ and $y$ represent the energy dissipation radio model used. Clearly E2 < E1 but in this case lot of uncompressed data is collected at the base station. Uncompressed data at base station not a problem as at base station power & space is not a constraint. When E1 > E2, it is not an optimal choice for saving energy and transmit data and we can avoid transmission for particular time as sensor will go to the sleep mode for three rounds. For these three rounds if node itself become cluster head then it can transmit the data otherwise after three rounds sensor node sends data directly to the near cluster head where E1 < E2. When E1 < E2 then a node will become a cluster head or a member of cluster head. If sensor node for E1 < E2 not become a cluster head nor become a member of cluster head for maximum three rounds then sensor node transmit the data directly to base station.

For E1 < E2 Sensor nodes send data to near cluster head and cluster head compress all the data received from its sensor members. To reduce the energy at cluster head level, it will not send data directly to base station it will follows some steps:

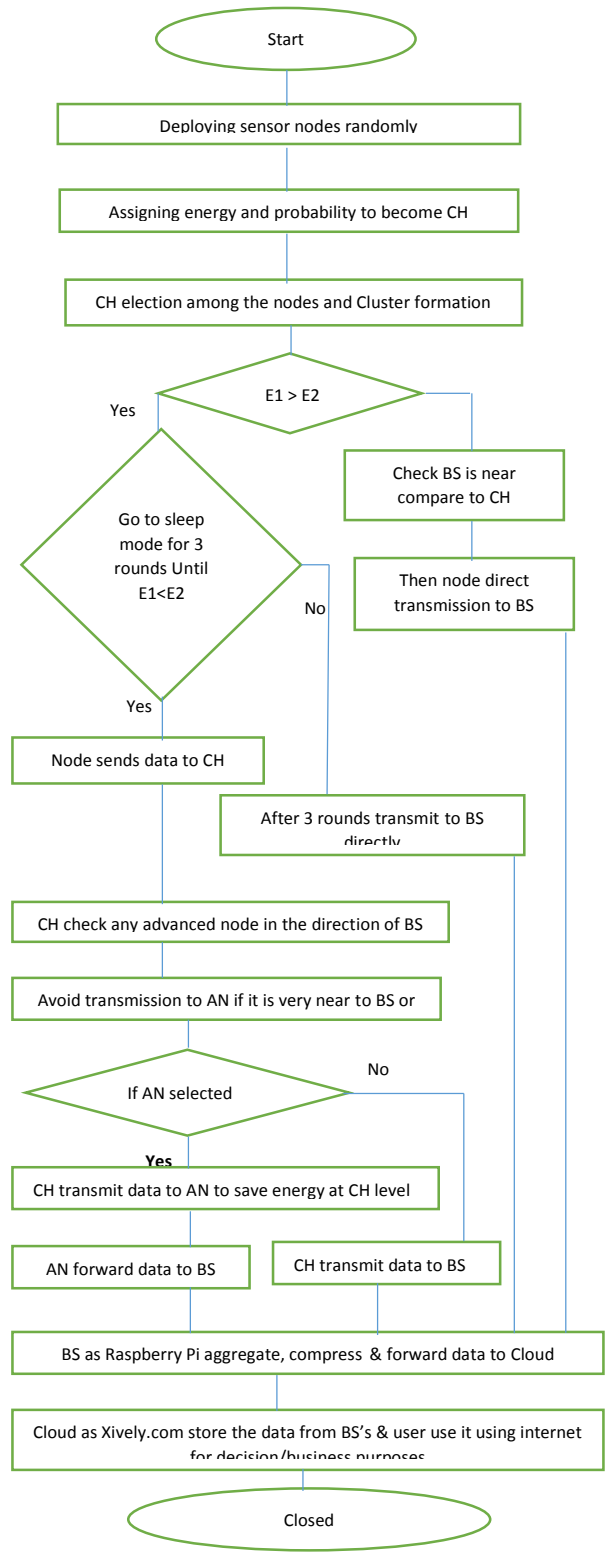

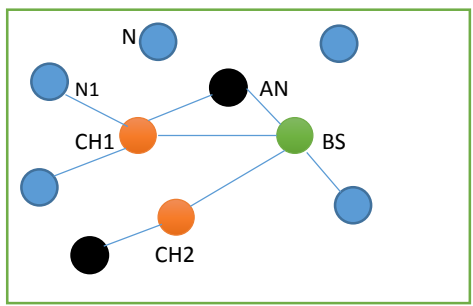

*Figure 2 Cluster head formation & selection*

*Fig. 3 Flow Chart of proposed Algorithm*

- If there is an advanced node between cluster head and base station then cluster head send data to advanced node then advanced node forward it to base station.
- If AN is very near to BS then cluster head first send data to AN which is not a good choice because AN first receive the data then forward it to the base station hence total energy consumption (receiving and sending) more compare to directly transmission of data from CH to BS.
- If AN is very near to CH then first AN receive all the data and then forward it to BS hence more energy consumption at AN (as long transmission will consume more energy). Delay is another parameter for QOS, if data transmission happens directly from CH to BS then minimum delay occur otherwise delay will increased.
- For best result AN should be exact between CH and BS. Hence uniform consumption of energy at AN, CH, and minimum delay will occur.

The performance of the WSN ca be evaluated on the basis of the following metrics and radio model in table 2 used for EMEEDP.

Stability Period is the time interval between network start until the death of the first node. Network lifetime is the time interval between network start until the death of the last node. Number of cluster heads per round will reflect the number of cluster heads formed in each round. Number of Alive nodes per round will measure the total number of live nodes per round. Throughput will measure the total number of packets which are sent to base station.

| Parameter | Value |
|---|---|
| BS Location | (50,50) |
| Energy consumed in the electronics circuit to transmit or receive the message, Eelec | 50 nJ/bit |
| Energy consumed by the amplifier to transmit message at a short distance, Efs | 10 pJ/bit/m^2 |
| Energy consumed by the amplifier to transmit message at a longer distance, Emp | 0.0013 pJ/bit/m^4 |
| Initial Energy, E0 | 0.5 J |
| Data Aggregation Energy, EDA | 5 nJ/bit/message |
| Message Size | 4000 bits |
| Popt | 0.1 |
| Cluster Radius | 25 m |
| Threshold distance, d0 | 70m |

*Table 2 Radio parameters for EMEEDP*

## VI. CONCLUSION & FUTURE WORK

EMEEDP is a multi-hop energy efficient clustering algorithm for heterogeneous WSN. The proposed EMEEDP is an extension of MEEP and it takes the full advantage of computation and power heterogeneity. It improves the CH lifetime resulting increased network lifetime, stability period, throughput and reduce delay of WSN network. Clearly if CH select AN as an intermediary node then CH will save the transmission energy because transmission energy proportional to the distance. If AN near to BS or CH then only delay will increase. Implemented a WSN to show the working protocols and multi-path communication using Wi-Fi technology. In future work try to implement WSN using ZigBee stack and WiMAX technology to provide heterogeneity in communication.

For future work, EMEEDP can be extended to deal with an energy efficient algorithm through data aggregation in a mobile sensor network.